\documentclass[prb,twocolumn,showpacs,showkeys,preprintnumbers,,superscriptaddress,citeautoscript,amsmath,amssymb]{revtex4}
\usepackage{mathrsfs}
\usepackage{amssymb,amsmath}
\usepackage{graphics}
\usepackage{bm}
\usepackage{epsfig}
\usepackage{color}
\usepackage{amsfonts}

\usepackage{amscd}

\begin{document}

\title{Three-terminal thermoelectric transport through a molecular junction}

\author{O. Entin-Wohlman}
\email{oraentin@bgu.ac.il}

\altaffiliation{Also at Tel Aviv University, Tel Aviv 69978,
Israel}

\affiliation{Department of Physics and the Ilse Katz Center for
Meso- and Nano-Scale Science and Technology, Ben Gurion
University, Beer Sheva 84105, Israel}

\affiliation{Albert Einstein Minerva Center for Theoretical
Physics, Weizmann Institute of Science, Rehovot 76100, Israel}

\author{Y. Imry}

\affiliation{Department of Condensed Matter Physics,  Weizmann
Institute of Science, Rehovot 76100, Israel}

\author{A. Aharony}

\altaffiliation{Also at Tel Aviv University,
Tel Aviv 69978, Israel}

\affiliation{Department of Physics and the Ilse Katz Center for
Meso- and Nano-Scale Science and Technology, Ben Gurion
University, Beer Sheva 84105, Israel}

\date{\today}

\begin{abstract}

The thermoelectric transport through a molecular bridge is discussed, with an emphasis on the effects of inelastic processes of the transport electrons caused by the coupling to the vibrational modes of the molecule. In particular it is found that when the molecule is strongly coupled to a thermal bath of its own, which may be at a temperature different from those of the electronic reservoirs, a heat current between the molecule and the electrons can be converted into an electric current. Expressions for the transport coefficients governing this conversion and similar ones are derived, and a possible scenario for increasing their magnitudes is outlined.

\end{abstract}


\maketitle

\section{Introduction}

\label{INTRO}

The investigation of thermoelectric phenomena   in nanoscale devices   at low temperatures has several interesting aspects.  From the practical point of view,  it is important to understand the heat flow and the dissipation because the  heat generated by  electric potentials used to switch-on transport currents inevitably  induces decoherence in the quantum functioning  of the device and also leads to dissipation. In bulk conductors, thermoelectric transport necessitates an asymmetry between holes and electrons, which is usually small.   In mesoscopic structures this asymmetry may be fairly high and can be also controlled experimentally.
One would hence like to have a full picture of the  symmetries and the inter-relations dominating the various transport  coefficients of a small mesoscopic system, in particular the effects of inelastic processes. Indeed, when transport is through a molecular bridge, the tunneling  electrons may undergo inelastic collisions with the vibrational modes even in the linear-response regime. This is because at finite temperatures  the transport electrons  may excite or de-excite the phonons and thus exchange energy with them.
These inelastic processes modify the electronic  transport coefficients,  leading to  the question of what, if any,  are  the analogues of the (bulk)  Onsager-Casimir relations.
Another intriguing issue is the possibility to convert heat from the vibrations into an electric current between the electronic reservoirs, or vice-versa.

Early studies of thermoelectric  transport coefficients of microstructures were based on the Landauer approach \cite{SIVAN,STREDA,BUTCHER,PROETTO} which was also extended to include mesoscopic superconductors. \cite{CLAUGHTON}
Once feasibility of measuring thermal and thermoelectric transport in atomic-scale samples had been established, \cite{PELTIER}
mesoscopic thermoelectric phenomena, e.g., peaks in the thermopower of a point contact orchestrated with the transitions among plateaux of  the quantized conductance, \cite{MOLENKAMP} or oscillations (as a function of a gate voltage) in the same coefficient measured on a quantum dot \cite{QDTP,PEPPER,LUDOPH},
were detected and analyzed. \cite{STREDA,CBEENAKKER}
The thermopower measured on nanotubes
was found to be unexpectedly high, and this  was attributed to a broken electron-hole symmetry. \cite{COHEN,PEREZ} Similarly, nanotubes exhibited  enhanced thermal conductivity, \cite{MCEUEN} as did also silicon nanowires. \cite{HOCHBAUM}
The dependence of the thermoelectric response on the length of the atomic chain has been recently computed within density-functional theory. \cite{CUEVAS}
Being based on the Landauer approach, the above-mentioned  theoretical studies mainly focused
on elastic processes of the transport electrons. Later on,  effects of inelastic electron-electron processes and electronic correlations (increasingly important at lower temperatures), as well as that of an applied magnetic field,  on the thermopower produced in
large \cite{MATVEEV} and single-level  \cite{KIM} quantum dots,  and also in  quantum wires \cite{flensberg} were considered. The effect of attractive electronic interactions on the thermopower was considered in Ref. ~\onlinecite{Karen}.

Inelasticity of electronic processes should  play a significant role
in thermoelectric transport through molecular bridges, also in the nonlinear regime. \cite{FLEN}
Indeed,  a density-functional  computation of the nonlinear differential conductance of gold wires attributed changes in the I-V characteristics to phonon heating, \cite{FREDERIKSEN,NITZAN}   and the thermopower coefficient was proposed as a tool to monitor the  excitation spectrum of a molecule forming the junction between two leads. \cite{ERAN,FINCH}
It was  suggested that the Seebeck effect in such bridges can be used for converting heat into electric energy \cite{MURPHY}, and  to determine the location of the Fermi level of the transport electrons relative
to the molecular levels, and also the sign of the dominant charge carriers, either for a molecular conductor, \cite{PAULSSON,TAIWAN,baranger} or for an atomic chain. \cite{ZHENG,DVIRA} This  was confirmed experimentally: the Seebeck coefficient
as measured by STM yielded that in the case of the benzenedithiol family sandwiched between two gold electrodes the charge carriers are holes passing through the HOMO, whose location with respect to the metal Fermi level was determined from the magnitude of the coefficient. \cite{Reddy}

Inelastic electron-vibration interactions are not included in several of the theoretical studies devoted to molecular junctions (see, e.g., Refs. ~\onlinecite{CUEVAS}, ~\onlinecite{PAULSSON}, and ~\onlinecite{ZHENG})
or are treated at off-resonance conditions, expanding them in the molecular-lead coupling. \cite{ERAN} When these interactions are ignored, the transport coefficients have the same functional form as in bulk conductors, with the energy-dependent transmission coefficient and its derivative  replacing  the conductivity. \cite{CUEVAS}
Notwithstanding the relative smallness, often,  of the inelastic corrections to the thermoelectric transport, their study is still of interest because of  fundamental questions related to the symmetries of the conventional transport coefficients, and since they give rise to additional coefficients connecting the heat transport
in-between the electrons and the phonons.

\begin{figure}[ hbtp]
\includegraphics[width=7cm]{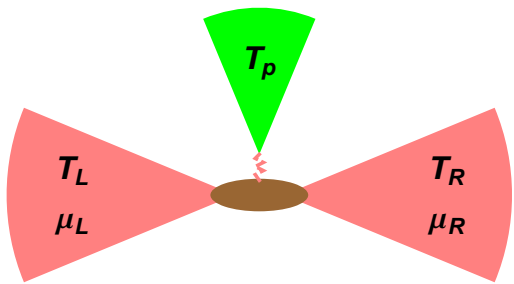}
\caption{(color online) A three-terminal system, modeled by a resonant level attached to two electronic reservoirs, having different chemical potentials and temperatures $\mu_{\rm{L,R}}$ and $T_{\rm{L,R}}$, respectively. An electron residing on the level interacts with its vibrational modes. The population of these phonons can be determined by the transport electrons (a ``floating molecule") or by a coupling to a phonon source kept at temperature $T_{\rm P}$. }\label{sys}
\end{figure}

Here we study the heat and charge transport in a small mesoscopic (or nanometric) system depicted schematically in Fig. \ref{sys}: a molecule attached to two electronic (no phonons) reservoirs, 
held in general at different temperatures, $T_{\rm L,R}$, and at different chemical potentials, $\mu_{\rm L,R}$. We distinguish between two (extreme) situations. In the first, the molecule is ``floating" and is attached solely to the leads; then the vibration population is determined by the transport electrons alone. In that case, the system is a two-terminal junction.
In the second case, the molecule is coupled to its own (typically a phonon-) heat bath which is kept at the temperature $T_{\rm P}$,   making the system a three-terminal one. It is then  assumed implicitly   that the coupling of the molecule to that heat bath largely exceeds its coupling to the transport electrons. The latter is determined by our small parameter, the coupling between the molecule vibrations and the transport electrons,  $\gamma$. Thus, we assume that the relaxation time due to the coupling to the heat bath,  $\tau_{\rm P}$, is short on the scale        $ \Gamma^{2}/(\gamma^{2}\omega_{0})$       [see Eq. (\ref{phe}) in  appendix \ref{DETCAL}],  $\Gamma$ being the level width on the molecule, due to the coupling with the leads  and $\omega_{0}$ is the frequency of the vibrations. $\hbar / \tau_{\rm P}$ may still be very small on all other physical scales, such as
$\hbar\omega_{0}$  and  $\Gamma$. The phonon bath may be realized simply by an electronically insulating hard substrate (assuming that the large Kapitsa-type phonon thermal resistance between the lead and the sample is large enough to sufficiently reduce the thermal contact of the molecule to the substrate via the leads), or a piece of such material touching the junction, each of those held at  a temperature $T_{\rm P}$. A vacuum gap between the two {\it  separate} substrates for the two leads would be ideal. However, with present fabrication technology, this appears possible for a quantum dot but not for a small molecule.

The consideration of  the entropy production of such a three-terminal system is quite illuminating. Using the thermodynamic identity
$TdS=dE-\mu dN$, \cite{COM1}
one finds that the dissipation at the left (right) reservoir leads to
\begin{align}
\dot{S}^{}_{\rm L(R)}=\frac{1}{ T_{\rm L(R)}}\Bigl (\dot{E}^{}_{\rm L(R)}-\mu^{}_{\rm L(R)}\dot{N}^{}_{\rm L(R)}\Bigr )\ .\label{SDL}
\end{align}
Here, $-\dot{E}_{\rm L(R)}$ is energy current emerging from the left (right) reservoir, while $-\dot{N}_{\rm L(R)}$ is the particle current leaving the left (right) reservoir.
Adding to Eqs. (\ref{SDL})  the entropy production of the phonon heat bath, $\dot{S}_{\rm P}=\dot{E}_{\rm P}/T_{\rm P}$, where $-\dot{E}_{\rm P}$ is the energy current leaving that bath,  yields the total dissipation of the system,
\begin{align}
\dot{S}^{}_{\rm P}&+\dot{S}^{}_{\rm L}+\dot{S}^{}_{\rm R}=\frac{\dot{E}_{\rm P}}{ T_{\rm P}}\nonumber\\
&+\frac{1}{ T_{\rm L}}\Bigl (\dot{E}^{}_{\rm L}-\mu^{}_{\rm L}\dot{N}^{}_{\rm L}\Bigr )+\frac{1}{T_{\rm R}}\Bigl (\dot{E}^{}_{\rm R}-\mu^{}_{\rm R}\dot{N}^{}_{\rm R}\Bigr )\ .\label{TSD}
\end{align}
Charge conservation implies that
\begin{align}
\dot{N}^{}_{\rm L}+\dot{N}^{}_{\rm R}=0\ ,\label{CCon}
\end{align}
while energy conservation requires
\begin{align}
\dot{E}^{}_{\rm L}+\dot{E}^{}_{\rm R}+\dot{E}^{}_{\rm P}=0\ .\label{ECon}
\end{align}
In the linear-response regime
all three temperatures (see Fig. \ref{sys}) are only slightly different,
\begin{align}
T^{}_{\rm L(R)}&=T\pm\frac{\Delta T}{2}\ ,\nonumber\\
T^{}_{\rm P}&=T+\Delta T^{}_{\rm P}\ ,\label{LRT}
\end{align}
and  the chemical potentials differ by a small amount,
\begin{align}
\mu_{\rm L(R)}=\mu\pm\frac{\Delta\mu}{2}\ .\label{LRM}
\end{align}
Expanding Eq. (\ref{TSD}) and using Eqs. (\ref{CCon}) and (\ref{ECon}) yields
\begin{align}
\dot{S}^{}_{\rm P}&+\dot{S}^{}_{\rm L}+\dot{S}^{}_{\rm R}=\frac{\Delta T^{}_{\rm P}}{T^{2}}(-\dot{E}^{}_{\rm P})
+\frac{\Delta\mu /e}{T}I+\frac{\Delta T}{T^{2}}I^{}_{\rm Q}\ ,
\end{align}
where $I$ is the net charge current flowing from the left reservoir to the right one,
\begin{align}
I=-\frac{e}{2}\Bigl (\dot{N}^{}_{\rm L}-\dot{N}^{}_{\rm R}\Bigr )\ ,\label{I}
\end{align}
while $I_{\rm Q}$ is the net heat current carried by the electrons,
\begin{align}
I^{}_{\rm Q}=I^{}_{\rm E}-(\mu/e)I\ ,\ {\rm with}\ \ I^{}_{\rm E}=-\frac{1}{2}\Bigl (\dot{E}^{}_{\rm L}-\dot{E}^{}_{\rm R}\Bigr )\ . \label{IQ}
\end{align}
Finally,  the heat current flowing from the phonon bath to the quantum system is simply given from the condition of energy conservation, \begin{align}
-\dot{E}^{}_{\rm P}=\dot{E}^{}_{\rm L}+\dot{E}^{}_{\rm R}\  . \label{IQP}
\end{align}
Thus, the entropy production of our three-terminal system is a simple example of the general expressions for linear transport, consistent with the Onsager theory. \cite{LL}

Since our molecular bridge is not necessarily at equilibrium
within the transport process,
it exchanges energy with the phonons of the phonon reservoir by going up/down in the vibrational
ladder with absorbing/emitting a phonon in the bath. This is the physical origin of  the current $-\dot{E}_{\rm P}$.
On the other hand, when the molecule is floating, then
$-\dot{E}_{\rm P}$ vanishes. Since  $-\dot{E}_{\rm P}$ is proportional to the rate of change of the vibrational level population on the dot  (see Appendix \ref{DETCAL} for details) this in turn will determine the vibration population that will
adjust itself according to the temperature and chemical potential differences applied to the electrons. In this situation our device becomes a two-temrinal one, and the energy current carried by the electrons is conserved.

In Sec. \ref{CURS} we outline our model, and give explicit expressions for all three currents $I$, $I_{\rm Q}$, and $-\dot{E}_{\rm P}$, and in Sec.
\ref{LINRES} we discuss them in the linear-response regime. In particular, we find there that by the three-terminal junction, one may convert the heat current from the phonon bath into  electric and heat currents carried by the electrons even at zero bias voltage
and when $T_{\rm L}=T_{\rm R}$.  In Sec. \ref{2t} we discuss the necessary conditions for this  conversion to be established, i.e.,  the junction couplings to the electronic reservoirs  should not be spatially-symmetric and should not depend on the energy in an identical manner. We show that an opposite dependence on energy of the couplings to two electron reservoirs will tend to maximize the new transport coefficients we find.

\section{The  currents}

\label{CURS}

In our analysis, the molecular bridge is replaced by a single resonant level; when a transport electron resides on the level, it interacts (linearly) with the phonons. Such a model, which neglects effects of spin and electronic correlations, is applicable in the Coulomb blockade regime for low-energy molecular levels. (We also ignore the possibility of the Kondo effect to develop, namely the average temperature of the system should exceed the Kondo temperature.) The
model Hamiltonian (see Fig. \ref{sys}) is thus
\begin{align}
{\cal H}={\cal H}^{}_{\rm L}+{\cal H}^{}_{\rm R}+{\cal H}^{}_{\rm dot}+{\cal H}^{}_{\rm coup}\ ,\label{FH}
\end{align}
in which ${\cal H}_{\rm L(R)}$ is the  Hamiltonian of the left (right) lead,
\begin{align}
{\cal H}^{}_{\rm L(R)}=&\sum_{k(p)}\epsilon^{}_{k(p)}c^{\dagger}_{k(p)}c^{}_{k(p)}
\  ,\label{HL}
\end{align}
 [using $k(p)$ for the left (right) lead].  The Hamiltonian of the bridge, which includes the electron-phonon interaction,  is
\begin{align}
{\cal H}^{}_{\rm dot}=\epsilon^{}_{0}c^{\dagger}_{0}c^{}_{0}+\omega^{}_{0}( b^{\dagger}b+\frac{1}{2})+\gamma (b+b^{\dagger})c^{\dagger}_{0}c^{}_{0}\ , \label{HD}
\end{align}
where $\omega_{0}$ is the frequency of the harmonic oscillator, and $\gamma$ is its coupling to the transport  electrons.
(We use units in which $\hbar=1$.)
Finally, the coupling between the dot and the leads  is described  by
\begin{align}
{\cal H}^{}_{\rm coup}=\sum_{k}(V^{}_{k}c^{\dagger}_{k}c^{}_{0}+{\rm hc})+\sum_{p}(V^{}_{p}c^{\dagger}_{p}c^{}_{0}+{\rm hc})\ .
\end{align}
The operators $c^{\dagger}_{0}$, $c^{\dagger}_{k}$, and $c^{\dagger}_{p}$ ($c^{}_{0}$, $c^{}_{k}$, and $c^{}_{p}$)
create (destroy) an electron on the dot, on the left lead, and on the right lead, respectively, while $b^{\dagger}$ ($b$) creates (destroys) an excitation of the harmonic oscillator, of frequency $\omega_{0}$.
The  electron distributions of the leads, $f_{\rm L}$ and $f_{\rm R}$, are given  by
\begin{align}
f_{\rm L(R)}^{}(\omega )=\Bigl (1+\exp [\beta^{}_{\rm L(R)}(\omega -\mu^{}_{\rm L(R)})]\Bigr )^{-1}\ , \label{FLR}
\end{align}
where $\beta_{\rm L(R)}=1/(k_{\rm B}T_{\rm L(R)})$.

The couplings of the leads to the resonance level broadens it, such that
\begin{align}
\Gamma^{}_{\rm L(R)}(\omega )=2\pi\sum_{k(p)}|V^{}_{k(p)}|^{2}_{}\delta (\omega -\epsilon^{}_{k(p)})\ ,\label{PARLR}
\end{align}
are the partial widths brought about by the left and the right leads. These couplings are treated to all orders, encompassing  the case in which the transport electrons excite effectively the phonons (the  dwell time of the electrons on the junction largely exceeds the response time of the oscillator,  about $\omega_{0}^{-1}$), and also the inverse situation.
Strictly speaking, the Hamiltonian (\ref{FH}) pertains to a ``floating" molecule, which is not coupled to a heat bath of its own; however, the analysis presented above in Sec. \ref{INTRO} enables us to consider the three-terminal case (see Fig. \ref{sys}) as well.

The explicit calculation of the currents is carried out up to second order in the electron-phonon coupling,  using the Keldysh technique, \cite{HAUG} and the details are given in Appendix \ref{DETCAL}.
We find that the charge current consists of two terms, which can be related to elastic and inelastic transitions of the electrons through the junction  [the first and the second terms in Eq. (\ref{Ic}), respectively]
\begin{align}
&I=e\int\frac{d\omega}{2\pi}| G^{r}_{00}(\omega )|^{2}_{}\Gamma^{}_{\rm L}(\omega )\Gamma^{}_{\rm R}(\omega )\Bigl (f^{}_{\rm L}(\omega )-f^{}_{\rm R}(\omega )\Bigr )\nonumber\\
&+e\gamma^{2}\int\frac{d\omega}{2\pi}|{\cal G}^{r}_{00}(\omega^{}_{+} )|^{2}_{}|{\cal G}^{r}_{00}(\omega ^{}_{-} )|^{2}_{}\nonumber\\
&\times\Bigl (\Gamma^{}_{\rm R}(\omega ^{}_{+})\Gamma^{}_{\rm L}(\omega^{}_{-} )F^{}_{\rm RL}(\omega )-{\rm L}\leftrightarrow{\rm R}\Bigr )\ ,\label{Ic}
\end{align}
where we have introduced the abbreviations
\begin{align}
\omega^{}_{\pm}=\omega\pm\frac{\omega^{}_{0}}{2}\ .
\end{align}
Here, $G_{00}$ is the Green function of the dot, given by Eq. (\ref{DOTR}), and ${\cal G}_{00}$ is its counterpart when the coupling to the phonons is ignored, i.e.,
\begin{align}
|{\cal G}^{r}_{00}(\omega )|^{2}_{}=\Big |\frac{1}{\omega -\epsilon^{}_{0}+i\Gamma (\omega )/2{}}\Big |^{2}\  \label{NOPHO}
\end{align}
represents the bare Breit-Wigner resonance on the dot, with
\begin{align}
\Gamma (\omega )=\Gamma_{\rm L}(\omega )+\Gamma_{\rm R}(\omega )\ .\label{GAMA}
\end{align}
Finally,
\begin{align}
F^{}_{\alpha \alpha '}(\omega )&=N[1-f^{}_{\alpha}(\omega^{}_{+})]f^{}_{\alpha '}(\omega^{}_{-} )\nonumber\\
&
-[1+N][1-f^{}_{\alpha '}(\omega^{}_{-} )]f^{}_{\alpha}(\omega^{}_{+})\ \label{F}
\end{align}
embodies the populations of the electrons ($f_{\rm L,R}$) and the phonons ($N$). Note that the latter population is {\em not} necessarily given by the Bose-Einstein distribution; this is the case only when the molecule is strongly coupled to a heat bath of its own (this distribution is denoted below  by $N_{\rm T}$). In the case of the floating molecule, the population $N$ is determined by the transport electrons as explained below and in Appendix   \ref{DETCAL}     .

The energy current carried by the electrons, $I_{\rm E}$, [see Eq. (\ref{IQ})] is shown in Appendix \ref{DETCAL} to be
\begin{align}
&I^{}_{\rm E}=\int\frac{d\omega}{2\pi}| G^{r}_{00}(\omega )|^{2}_{}\omega\Gamma^{}_{\rm L}(\omega )\Gamma^{}_{\rm R}(\omega )\Bigl (f^{}_{\rm L}(\omega )-f^{}_{\rm R}(\omega )\Bigr )\nonumber\\
&+\gamma^{2}\int\frac{d\omega}{2\pi}|{\cal G}^{r}_{00}(\omega^{}_{+} )|^{2}_{}|{\cal G}^{r}_{00}(\omega^{}_{-} )|^{2}_{}\nonumber\\
&\times\Bigl (\frac{\omega^{}_{0}}{2}[\Gamma^{}_{\rm R}(\omega^{}_{+} )\Gamma^{}_{\rm R}(\omega ^{}_{-})F^{}_{\rm RR}(\omega )-({\rm R}\rightarrow {\rm L})]
\nonumber\\
&+\omega [\Gamma^{}_{\rm R}(\omega^{}_{+} )\Gamma^{}_{\rm L}(\omega ^{}_{-})F^{}_{\rm RL}(\omega )-({\rm L}\leftrightarrow {\rm R})]\Bigr )
\ , \label{NEDOT}
\end{align}
where again the first and second terms pertain to the elastic and inelastic contributions to the electronic energy current.
The energy current carried by the phonons [see Eqs. (\ref{ECon}) and (\ref{IQP})] is
\begin{align}
-\dot{E}_{\rm P}=&\int\frac{d\omega}{2\pi}|{\cal G}^{r}_{00}(\omega ^{}_{+})|^{2}
|{\cal G}^{r}_{00}(\omega ^{}_{-})|^{2}\nonumber\\
&\times \gamma^{2}\omega^{}_{0}\sum_{\alpha,\alpha '={\rm L,R}}\Gamma^{}_{\alpha}(\omega^{}_{+} )\Gamma^{}_{\alpha '}(\omega ^{}_{-})F^{}_{\alpha\alpha '}(\omega )\ .\label{DOTEP}
\end{align}
In the next section we examine these currents in the linear-response regime.

\section{The linear-response regime}

\label{LINRES}

The temperatures and the chemical potentials of the three-terminal junction are given by Eqs. (\ref{LRT}) and (\ref{LRM}). In the linear-response regime, one expands the   currents [see Eqs. (\ref{I}), (\ref{IQ}), and (\ref{IQP})]  to first order in $\Delta \mu$, $\Delta T$, and $\Delta T_{\rm P}$.
In order to express the resulting  transport coefficients in a convenient form, we note that all integrals resulting from the elastic processes include  the function
\begin{align}
F_{}^{\rm el}(\omega )=\beta f(\omega )[1-f(\omega )]|G^{r}_{00}(\omega )|^{2}\ ,\label{FEL}
\end{align}
where $f(\omega )$ is the thermal-equilibrium Fermi distribution of temperature $T$, and  $\beta=1/(k_{\rm B}T)$. The transport coefficients coming from the inelastic processes include in their integral forms the function
\begin{align}
F_{}^{\rm inel}(\omega )&=\gamma^{2}|{\cal G}^{r}_{00}(\omega^{}_{+})|^{2}|{\cal G}^{r}_{00}(\omega^{}_{-})|^{2}\nonumber\\
&\times N^{}_{\rm T}\beta f(\omega^{}_{-})[1-f(\omega^{}_{+})]
\ , \label{FINEL}
\end{align}
where $N_{\rm T}$ is the thermal-equilibrium Bose distribution function of temperature $T$.

The relations between the currents and the driving forces in the linear-response regime can be written in the
matrix form
\begin{align}
\left [\begin{array}{c}I\\ I^{}_{\rm Q} \\-\dot{E}^{}_{\rm P}\end{array}\right ]
={\cal M}
\left [\begin{array}{c}\Delta\mu/e\\  \Delta T/T \\ \Delta T^{}_{\rm P}/T\end{array}\right ]\ ,\label{MAT}
\end{align}
where the matrix of the transport coefficients, ${\cal M}$, is
\begin{align}
{\cal M}=\left [\begin{array}{ccc}{\rm G}\ \ & \ \  {\rm K} \ \ & \ \ {\rm X}^{\rm P}_{}\\
 {\rm K}\ \ &\ {\rm K}^{}_{2}+{\rm K}^{\rm P}_{2}\  & \ \ \widetilde{\rm X}^{\rm P}_{}\\
 {\rm X}^{\rm P}\ \  & \ \ \widetilde{\rm X}^{\rm P}_{}\ \ &\ \ {\rm C}_{}^{\rm P}\end{array}\right ]\ .\label{M}
  \end{align}
Let us first describe the conventional transport coefficients, pertaining to the transport by the electrons.
 In Eq. (\ref{M}),  G is the electrical conductance,
\begin{align}
{\rm G}={\rm G}_{}^{\rm el}+{\rm G}_{}^{\rm inel}\ ,
\end{align}
which consists of the contribution of elastic processes,
\begin{align}
{\rm G}_{}^{\rm el}&=\frac{e^{2}}{2\pi}\int d\omega F_{}^{\rm el}(\omega )\Gamma^{}_{\rm L}(\omega )\Gamma^{}_{\rm R}(\omega ) ,
\end{align}
and the contribution of the inelastic ones
\begin{align}
{\rm G}_{}^{\rm inel}=&\frac{e^{2}}{2\pi}\int d\omega F_{}^{\rm inel}(\omega )\nonumber\\
&\times\Bigl (\Gamma^{}_{\rm L}(\omega^{}_{+} )\Gamma^{}_{\rm R}(\omega^{}_{-} )
+\Gamma^{}_{\rm L}(\omega^{}_{-} )\Gamma^{}_{\rm R}(\omega^{}_{+} )\Bigr )\ . \label{Gin}
\end{align}
Clearly, Eq. (\ref{Gin})   corresponds to the two inelastic processes by which the transport electron excites or de-excites the phonon  upon moving between the reservoirs. The transport coefficient yielding the thermopower and the Seebeck effect, K, and the one giving
the main contribution to the electric thermal conductance, K$_{2}$,  also consist of two contributions each,
\begin{align}
{\rm K}&={\rm K}_{}^{\rm el}+{\rm K}_{}^{\rm inel}\ ,\nonumber\\
{\rm K}^{}_{2}&={\rm K}_{2}^{\rm el}+{\rm K}_{2}^{\rm inel}\ ,
\end{align}
with
\begin{align}
{\rm K}_{}^{\rm el}&=\frac{e^{}}{2\pi}\int d\omega F_{}^{\rm el}(\omega )
(\omega -\mu )\Gamma^{}_{\rm L}(\omega )\Gamma^{}_{\rm R}(\omega ) ,\nonumber\\
{\rm K}_{}^{\rm inel}&=\frac{e^{}}{2\pi}\int d\omega F_{}^{\rm inel}(\omega )
(\omega -\mu )\nonumber\\
&\times\Bigl (\Gamma^{}_{\rm L}(\omega^{}_{+} )\Gamma^{}_{\rm R}(\omega^{}_{-} )+
\Gamma^{}_{\rm L}(\omega^{}_{-} )\Gamma^{}_{\rm R}(\omega^{}_{+} )\Bigr ) \ , \label{K}
\end{align}
and
\begin{align}
{\rm K}_{2}^{\rm el}&=\frac{1^{}}{2\pi}\int d\omega F_{}^{\rm el}(\omega )
(\omega -\mu )^{2}\Gamma^{}_{\rm L}(\omega )\Gamma^{}_{\rm R}(\omega ) ,\nonumber\\
{\rm K}_{2}^{\rm inel}&=\frac{1^{}}{2\pi}\int d\omega F_{}^{\rm inel}(\omega )
(\omega -\mu )^{2}\nonumber\\
&\times\Bigl (\Gamma^{}_{\rm L}(\omega^{}_{+} )\Gamma^{}_{\rm R}(\omega^{}_{-} )+
\Gamma^{}_{\rm L}(\omega^{}_{-} )\Gamma^{}_{\rm R}(\omega^{}_{+} )\Bigr ) \ . \label{K2}
\end{align}
All other coefficients appearing in Eq. (\ref{M}) result from the inelastic processes. One of them,  ${\rm K}^{\rm P}_{2}$, just augments the (conventional) ratio K$_{2}$ between the heat current carried by the electrons and the temperature gradient $\Delta T$ across the junction,
\begin{align}
{\rm K}^{\rm P}_{2}&=\frac{\omega^{2}_{0}}{8\pi}\int d\omega F^{\rm inel}_{}(\omega )\nonumber\\
&\times\Bigl (\Gamma^{}_{\rm R}(\omega^{}_{+})\Gamma^{}_{\rm L}(\omega^{}_{-})+
\Gamma^{}_{\rm L}(\omega^{}_{+})\Gamma^{}_{\rm R}(\omega^{}_{-})\Bigr )\ . \label{K2P}
\end{align}
It therefore follows that the electron-phonon interaction just renormalizes slightly the conventional transport coefficients of the  two-terminal single dot junction, but does
not lead to novel effects (see also Sec. \ref{2t} below).

On the other hand, keeping the phonon bath to which the molecule is attached at a  temperature
different from those of the electron reservoirs leads to new thermoelectric effects.
We find that there is an electric current flowing in response to the temperature difference $\Delta T_{\rm P}$ with the phonon bath, with the novel transport coefficient
\begin{align}
{\rm X}^{\rm P}_{}=&\frac{e\omega^{}_{0}}{2\pi}\int d\omega F^{\rm inel}(\omega )\nonumber\\
&\times\Bigl (\Gamma^{}_{\rm R}(\omega^{}_{+})\Gamma^{}_{\rm L}(\omega ^{}_{-})
-
\Gamma^{}_{\rm L}(\omega^{}_{+})\Gamma^{}_{\rm R}(\omega ^{}_{-})\Bigr )\ .\label{SP}
\end{align}
The same coefficient controls the heat current between the junction and the phonon bath in response to the chemical potential difference between the electronic reservoirs. Likewise, there is a heat current flowing between the electronic reservoirs in response to $\Delta T_{\rm P}$, which is governed by a coefficient analogous to Eq. (\ref{SP}),
\begin{widetext}
\begin{align}
\widetilde{\rm X}^{\rm P}_{}=
&\frac{\omega^{}_{0}}{2\pi}\int d\omega F^{\rm inel}(\omega )\Bigl [(\omega -\mu)\Bigl (\Gamma^{}_{\rm R}(\omega^{}_{+})\Gamma^{}_{\rm L}(\omega ^{}_{-})
-
\Gamma^{}_{\rm L}(\omega^{}_{+})\Gamma^{}_{\rm R}(\omega ^{}_{-})\Bigr )
+\frac{\omega^{}_{0}}{2}\Bigl (\Gamma^{}_{\rm R}(\omega^{}_{+})\Gamma^{}_{\rm R}(\omega ^{}_{-})
-
\Gamma^{}_{\rm L}(\omega^{}_{+})\Gamma^{}_{\rm L}(\omega ^{}_{-})\Bigr )\Bigr ], \label{SPQ}
\end{align}
\end{widetext}
with the same coefficient governing the heat current from the phonon reservoir in response to the electronic temperature difference $\Delta T$. Thus, the matrix of coefficients ${\cal M}$ obeys the Onsager symmetry relations also in the three-terminal situation with the two types of carriers and their interaction.

Finally, the coefficient ${\rm C}^{\rm P}$  gives the response of the heat current carried by the phonons to the temperature difference $\Delta T_{\rm P}$,
\begin{align}
C^{\rm P}_{}=\frac{\omega^{2}_{0}}{2\pi}\int d\omega F^{\rm inel}_{}(\omega )\Gamma^{}_{}(\omega^{}_{+})\Gamma^{}_{}(\omega ^{}_{-})\ ,
\end{align}
where we have used Eq. (\ref{GAMA}).

\section{Discussion}

\label{2t}

Using a simple model, we have considered the thermoelectric and thermal transport of electrons through a molecular bridge, in particular the subtle effects of the inelastic electron-vibrational mode processes. Of a paramount importance  is the mechanism by which the vibration population is determined.

When the molecule is not attached to any heat bath, the phonon population is determined by the voltage and the  temperature difference across the junction.
We show in Appendix \ref{DETCAL}
that in this case [see Eq. (\ref{N2})] the heat current  between the vibrations and the transport electrons is
\begin{align}
\dot{E}_{\rm P}^{}=\omega^{}_{0}\frac{dN}{dt}\ ,
\end{align}
where $N$ denotes the vibrational mode population. At steady-state that population does not vary with time, and consequently the heat current between the molecule and the junction vanishes. This requirement, in turn, fixes $\Delta T_{\rm P}$ in terms of $\Delta\mu$ and $\Delta T$, and consequently determines the vibration population [see Fig. \ref{sys} and Eqs. (\ref{LRT}) and (\ref{LRM})].
In other words,  the requirement that $-\dot{E}_{\rm P}=0$  yields
\begin{align}
{\rm X}^{\rm P}_{}\frac{\Delta\mu}{e}+\widetilde{\rm X}^{\rm P}_{}\frac{\Delta T}{T}=-{\rm C}^{\rm P}_{}\frac{\Delta T^{}_{\rm P}}{T}\ ,
\end{align}
and hence transforms the
three-terminal junction into a two-terminal one, with
\begin{align}
&\left [\begin{array}{c}I\\ I^{}_{\rm Q}\end{array}\right ]=\nonumber\\
&\left [\begin{array}{cc}{\rm G}-({\rm X}^{\rm P}_{})^{2}/{\rm C}^{\rm P}_{}&{\rm K}-{\rm X}^{\rm P}_{}\widetilde{\rm X}^{\rm P}_{}/{\rm C}^{\rm P}_{}\\
{\rm K}-{\rm X}^{\rm P}_{}\widetilde{\rm X}^{\rm P}_{}/{\rm C}^{\rm P}_{}&{\rm K}^{}_{2}+{\rm K}^{\rm P}_{2}-(\widetilde{\rm X}^{\rm P}_{})^{2}/{\rm C}^{\rm P}_{}\end{array}\right ]\left[\begin{array}{c}\Delta\mu /e \\ \Delta T/T\end{array}\right ]\ .
\end{align}
In this situation we find that the inelastic processes modify the transport coefficients, but do not give rise to any intriguing effects.

On the other hand, when the molecule is attached (strongly) to its own thermal bath, see Fig. \ref{sys}, such that the system becomes a three-terminal junction, the vibrational modes and the transport electrons may exchange heat, and a temperature difference between the phonons and the transport electrons can induce an electron current between the electronic reservoirs.  Likewise,
a voltage between the latter can induce a heat current to the phonons. These two new transport coefficients, having two types of carriers and including inelastic processes, are related by Onsager symmetry. This situation is characterized by the appearance of new transport coefficients that result solely from the inelastic transport processes [see Eqs.
(\ref{MAT}) and (\ref{M})], and requires the breaking of spatial symmetry between the two sides of the junction, $\Gamma_{\rm L}\neq\Gamma_{\rm R}$. Note in particular
the change of the relative sign of the combinations $\Gamma^{}_{\rm R}(\omega^{}_{+})\Gamma^{}_{\rm L}(\omega ^{}_{-})$ and
$\Gamma^{}_{\rm L}(\omega^{}_{+}) \Gamma^{}_{\rm R}(\omega ^{}_{-})$  between the expressions for the usual thermoelectric coefficients,
Eqs. (\ref{K}),  (\ref{K2}),   and (\ref{K2P}), and the new three-terminal ones, Eqs. (\ref{SP}) and (\ref{SPQ}). This change occurs because the latter expressions
are for the heat currents from each lead to the phonons and not between the two leads.
The analysis of the above combinations of the $\Gamma$'s can tell us how to maximize the new, three-terminal, thermoelectric coefficients. Usually the $\omega$-dependence of the resonance widths  $\Gamma$'s is not too strong. Let us then expand them around the  running $\omega$,
$$\Gamma^{}_{\rm L}(\omega') = \Gamma^{}_{\rm L}(\omega) + A^{}_{\rm L}( \omega' - \omega) + ...\ ,$$
with an analogous expansion  for $\Gamma^{}_{\rm R}$.
The crucial quantity is the one in parentheses on the right-hand side  of Eq. (\ref{SP}).
To  order $\omega_0$ it gives
\begin{align}
\omega_0 (A^{}_{\rm R}  \Gamma^{}_{\rm L}(\omega)  - A^{}_{\rm L} \Gamma^{}_{\rm R}(\omega) )\ . \label{crux}
\end{align}
To increase  the usual thermopower, we want the transmission to depend  strongly on energy.
In our case, to make the two terms in Eq. (\ref{crux}) add and not tend to cancel, we also want
$\Gamma^{}_{\rm L}$ and $\Gamma^{}_{\rm R}$ to have  {\em opposite} dependencies
on the frequency. One way to effect this is to have a lead with an electron-band material on the left lead, and
one with a hole-band material on the right lead. This will
however decrease the values of the usual two-terminal thermal and thermoelectric coefficients.
Hence, more down-to-earth estimates of the new thermoelectric coefficients require  realistic descriptions of the molecular bridge, which will depend on the type of molecules involved and other parameters of the system.

In order to elucidate the above considerations, we compute the coefficient governing the conversion of heat from the phonon bath into a voltage difference across the bridge,
\begin{align}
{\rm S}^{\rm P}_{}\equiv\frac{e{\rm X}^{\rm P}_{}}{T{\rm G}}\ ,\label{NeWT}
\end{align}
where ${\rm X}^{\rm P
}$ is given by Eq. (\ref{SP}). (This definition follows the conventional one for the thermopower.)
Let us  assume that the left reservoir is represented by an electron band, such that the partial width it causes to the resonant level is given by
\begin{align}
\Gamma^{}_{\rm L}(\omega )=\Gamma^{}_{\rm L}\sqrt{\frac{\omega -\omega^{}_{c}}{\omega^{}_{v}-\omega^{}_{c}}}\ ,\label{GAMLW}
\end{align}
while the right reservoir is modeled by a hole band, with
\begin{align}
\Gamma^{}_{\rm R}(\omega )=\Gamma^{}_{\rm R}\sqrt{\frac{\omega_{v} -\omega^{}_{}}{\omega^{}_{v}-\omega^{}_{c}}}\ .\label{GAMRW}
\end{align}
Here, $\omega_{c}$ is the bottom of the conductance band (on the left side of the junction), while $\omega_{v}$ is the ceiling of the
hole band (on the right one).  The energy integration determining the various transport coefficients is therefore limited to the region $\omega_{c}\leq\omega\leq\omega_{v}$.   (For convenience, we normalize the $\Gamma$'s by the full band width, $\omega_{v}-\omega_{c}$.)

\begin{figure}[ hbtp]
\includegraphics[width=7cm]{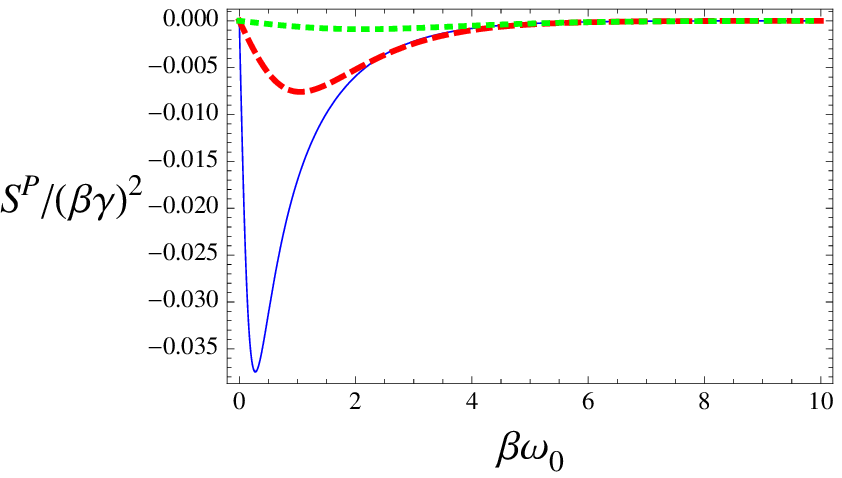}
\includegraphics[width=7cm]{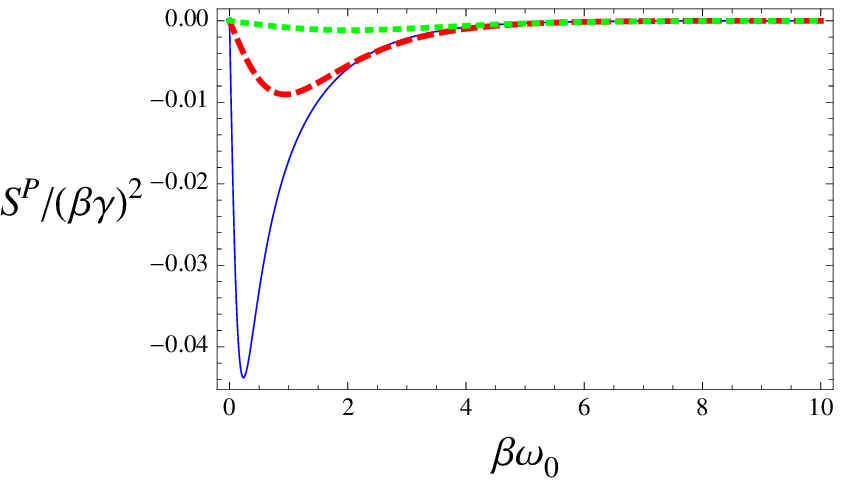}\caption{(color online) The coefficient ${\rm S}^{\rm P}$, Eq. (\ref{NeWT}), as a function of $\beta\omega_{0}$ for $\beta\Gamma_{\rm L}=0.2 $ (thin line),  $\beta\Gamma_{\rm L}=1 $ (dashed line),   and $\beta\Gamma_{\rm L}=5$ (dotted line). Upper panel, $\Gamma_{\rm R}=\Gamma_{\rm L}$, lower panel  $\Gamma_{\rm R}=0.7 \Gamma_{\rm L}$ [see Eqs. (\ref{GAMLW}) and (\ref{GAMRW})].  The total bandwidth is determined by $\beta\omega_{c}=-\beta\omega_{v}=100$.} \label{XPL}
\end{figure}

Measuring all energies appearing in the explicit expressions  in units  of the  temperature $\beta^{-1}$, and choosing $\mu=\epsilon_{0}$=0 for simplicity, we obtained the curves shown in Figs. \ref{XPL}.
One observes that the magnitude of the effect is non monotonic in the value of the vibration frequency $\omega_{0}$: being the outcome of inelastic processes, it vanishes at $\omega_{0}=0$, and also at $\beta\omega_{0}\gg1$, since then the vibrational level population becomes very small.  It  increases for values $\Gamma_{\rm L(R)}$ which are smaller than the temperature, and also increases when $\Gamma_{\rm L}\neq\Gamma_{\rm R}$, because then the electric conductance G becomes smaller.

\appendix

\section{Details of the currents' calculation}
\label{DETCAL}

The particle  and  the energy  currents can
be expressed in terms of the electronic Keldysh Green functions, in particular the Green function $G_{00}$ on the dot.
To this end we write the particle current emerging from the left (right) reservoir in the form
\begin{align}
\dot{N}^{}_{\rm L(R)}=\frac{d}{dt}\langle\sum_{k(p)}c^{\dagger}_{k(p)}c^{}_{k(p)}\rangle
=&\int\frac{d\omega}{2\pi}{\cal I}^{}_{\rm L(R)}(\omega )\ .\label{ILA}
\end{align}
Likewise, the energy current emerging from the left (right) reservoir can be shown to be given by
\begin{align}
\dot{E}^{}_{\rm L(R)}=
\frac{d}{dt}\langle\sum_{k(p)}\epsilon^{}_{k(p)}c^{\dagger}_{k(p)}c^{}_{k(p)}\rangle
=&\int\frac{d\omega}{2\pi}\omega{\cal I}^{}_{\rm L(R)}(\omega )\ .\label{ELA}
\end{align}
The Green function calculation yields
\begin{align}
{\cal I}^{}_{\rm L(R)}(\omega )&=-i\Gamma^{}_{L(R)}(\omega )\Bigl (G^{<}_{00}(\omega )\nonumber\\
&-f^{}_{\rm L(R)}(\omega )[G^{a}_{00}(\omega )-G^{r}_{00}(\omega )]\Bigr )\ ,\label{IINT}
\end{align}
where the superscripts $<$, $a$, and $r$, denote the lesser, advanced, and retarded Green function, respectively. The Fermi distributions, $f_{\rm L(R)}$,  are given in Eq. (\ref{FLR}), and the partial widths of the resonance level, $\Gamma_{\rm L(R)}$, in Eq. (\ref{PARLR}).

The Green function on the dot is calculated up to second order in the electron-phonon coupling $\gamma$. \cite{WE} One finds
\begin{align}
G^{<}_{00}(\omega )=G^{r}_{00}(\omega )\Bigl (\Sigma^{<}_{\rm P}(\omega )+\Sigma^{<}_{l}(\omega )\Bigr )G^{a}_{00}(\omega )\ ,\label{DOTKAT}
\end{align}
where
\begin{align}
G^{r}_{00}(\omega )=\Bigl (\omega -\epsilon^{}_{0}-\Sigma^{r}_{l}(\omega )-\delta\epsilon^{}_{\rm P} -\Sigma^{r}_{\rm P}(\omega )\Bigr )^{-1}\ .
\label{DOTR}
\end{align}
Here,
\begin{align}
\delta\epsilon^{}_{\rm P}&
=2i\frac{\gamma^{2} }{\omega^{}_{0}}\int\frac{d\omega}{2\pi} {\cal G}^{<}_{00}(\omega )\ , \label{SHIFT0}
\end{align}
is the polaron energy shift,
where
${\cal G}_{00}^{<}(\omega )$ is the lesser Green function on the dot in the absence of the coupling with the oscillator,
\begin{align}
{\cal G}^{<}_{00}(\omega )=i\frac{\Gamma^{}_{\rm L}(\omega )f^{}_{\rm L}(\omega )+
\Gamma^{}_{\rm R}(\omega )f^{}_{\rm R}(\omega )}{(\omega -\epsilon^{}_{0})^{2}+(\Gamma(\omega )/2)^{2}_{}}\ ,\label{GKK}
\end{align}
with $\Gamma =\Gamma_{\rm L}+\Gamma_{\rm R}$. Here we have ignored a possible shift in the resonance energy due to the coupling with the leads, since it is not expected to play a significant role.

As is seen from Eqs. (\ref{DOTKAT}) and  (\ref{DOTR}), the self energy on the dot includes two contributions. The first, $\Sigma_{l}$, is due to the coupling with the leads,
\begin{align}
\Sigma^{r}_{l}(\omega )&=-\frac{i}{2}\Bigl (\Gamma^{}_{\rm L}(\omega )+\Gamma^{}_{\rm R}(\omega )\Bigr )\ ,\nonumber\\
\Sigma^{<}_{l}(\omega )&=i \Bigl (\Gamma^{}_{\rm L}(\omega )f^{}_{\rm L}(\omega )+\Gamma^{}_{\rm R}(\omega )f^{}_{\rm R}(\omega )\Bigr )\ .
\end{align}
The second contribution to the self-energy results from the interaction with the phonons, and in second-order in $\gamma$ reads
\begin{align}
\Sigma_{\rm P}^{r}(\omega )
=&i\gamma^{2}\int\frac{d\omega '}{2\pi}\Bigl  (\frac{(1+N ){\cal G}^{>}_{00}(\omega ')-N{\cal G}^{<}_{00}(\omega ')}{\omega -\omega^{}_{0}-\omega '+ i 0^{+}}\nonumber\\
&+\frac{N{\cal G}^{>}_{00}(\omega ')-(1+N){\cal G}^{<}_{00}(\omega ')}{\omega +\omega^{}_{0}-\omega '+ i 0^{+}}\Bigr )\ ,\label{SIGHOr}
\end{align}
and
\begin{align}
&\Sigma^{<}_{\rm P}(\omega )=\gamma^{2}\Bigl (
 N{\cal G}^{<}_{00}(\omega -\omega^{}_{0})+(1+N) {\cal G}^{<}_{00}(\omega +\omega^{}_{0})\Bigr )\ ,\label{GSOFGn}
\end{align}
where $N$ denotes the phonon population. The lesser Green function ${\cal G}^{<}$ is given in Eq. (\ref{GKK}), and the greater one, ${\cal G}^{>}$, is given by the same expression with the distributions $f_{\rm L,R}$ replaced by $f_{\rm L,R}-1$.

Inserting the expressions for the Green function $G_{00}$ into Eq. (\ref{IINT}), one finds that ${\cal I}^{}_{\rm L(R)}(\omega )$ can be written as a sum of two terms, one arising from the elastic transitions of the transport electrons, and the other coming from the inelastic ones,
\begin{align}
{\cal I}^{}_{\rm L(R)}(\omega )={\cal I}^{\rm el}_{\rm L(R)}(\omega )+{\cal I}^{\rm inel}_{\rm L(R)}(\omega )\ .
\end{align}
The elastic-process contribution is
\begin{align}
{\cal I}^{\rm el}_{\rm L}(\omega )=|G^{r}_{00}(\omega )|^{2}\Gamma^{}_{\rm L}(\omega )\Gamma^{}_{\rm R}(\omega )[f^{}_{\rm R}(\omega )-f^{}_{\rm L}(\omega )]\ ,\label{LEL}
\end{align}
while the inelastic one is proportional to the strength of the  electron-phonon coupling,
\begin{widetext}
\begin{align}
{\cal I}^{\rm inel}_{\rm L}(\omega )=&\gamma^{2}\Gamma^{}_{\rm L}(\omega )|{\cal G}^{r}_{00}(\omega )|^{2}|{\cal G}^{r}_{00}(\omega -\omega^{}_{0})|^{2}\nonumber\\
&\times\sum_{\alpha ={\rm L,R}}\Gamma^{}_{\alpha}(\omega -\omega^{}_{0})[Nf^{}_{\alpha }(\omega -\omega^{}_{0})(1-f^{}_{\rm L}(\omega ))-(1+N)f^{}_{\rm L}(\omega )(1-f^{}_{\alpha }(\omega -\omega^{}_{0}))]\nonumber\\
&-\gamma^{2}\Gamma^{}_{\rm L}(\omega )|{\cal G}^{r}_{00}(\omega )|^{2}|{\cal G}^{r}_{00}(\omega +\omega^{}_{0})|^{2}\nonumber\\
&\times\sum_{\alpha ={\rm L,R}}\Gamma^{}_{\alpha}(\omega +\omega^{}_{0})[Nf^{}_{\rm L }(\omega ) (1-f^{}_{\alpha}(\omega+\omega^{}_{0}) )-(1+N)f^{}_{\alpha}(\omega +\omega^{}_{0})(1-f^{}_{\rm L }(\omega ))]\ ,\label{LIN}
\end{align}
\end{widetext}
where ${\cal G}^{r}_{00}$, the retarded Green function in the absence of the coupling to the vibrational modes,  is given in Eq. (\ref{NOPHO}). Since ${\cal I}_{\rm R}$ is obtained from Eqs. (\ref{LEL}) and (\ref{LIN}) upon interchanging L with R, it is easy to see that the elastic-process parts of both the particle and the energy currents are conserved (this is so because ${\cal I}^{\rm el}_{\rm L}(\omega )+{\cal I}^{\rm el}_{\rm R}(\omega )=0$).
The consideration of the inelastic-process part is a bit more delicate. By changing the integration variables one finds that
\begin{align}
&\int\frac{d\omega }{2\pi}\omega^{s}_{}{\cal I}^{\rm inel}_{\rm L}(\omega )
=\gamma^{2}\int\frac{d\omega}{2\pi}|{\cal G}^{r}_{00}(\omega ^{}_{+})|^{2}
|{\cal G}^{r}_{00}(\omega ^{}_{-})|^{2}\nonumber\\
&\times\Bigl (
\Gamma^{}_{\rm L}(\omega^{}_{+})\Gamma^{}_{\rm L}(\omega^{}_{-})(\omega^{s}_{+}-\omega^{s}_{-})F^{}_{\rm LL}(\omega )\nonumber\\
&+
\omega^{s}_{+}\Gamma^{}_{\rm L}(\omega^{}_{+})\Gamma^{}_{\rm R}(\omega^{}_{-})F^{}_{\rm LR}(\omega )\nonumber\\
&-
\omega^{s}_{-}\Gamma^{}_{\rm R}(\omega^{}_{+})\Gamma^{}_{\rm L}(\omega^{}_{-})F^{}_{\rm RL}(\omega )
\Bigr )\ ,\ \ \ s=0\ \ {\rm or}\ \ 1\ , \label{SEQ}
\end{align}
where $F_{\alpha \alpha '}$ is given in Eq. (\ref{F}).
Here, $\omega_{\pm}\equiv\omega\pm\omega_{0}/2$.
Hence, the inelastic-process parts of the particle current (for which $s=0$) are also conserved, i.e., $\int (d\omega /2\pi)[{\cal I}^{\rm inel}_{\rm L}(\omega )+
{\cal I}^{\rm inel}_{\rm R}(\omega )]=0$. Using Eqs. (\ref{ILA}), (\ref{LEL}), and  (\ref{SEQ}) in Eq. (\ref{I}) produces  Eq. (\ref{Ic}) for the charge current.

On the other hand, the energy current carried by the electrons alone is not conserved, since [using Eq. (\ref{SEQ}) with $s=1$]
\begin{align}
&\int\frac{d\omega }{2\pi}\omega^{}_{}[{\cal I}^{\rm inel}_{\rm L}(\omega )+
{\cal I}^{\rm inel}_{\rm R}(\omega )]=\omega^{}_{0}\gamma^{2}\int\frac{d\omega}{2\pi}|{\cal G}^{r}_{00}(\omega ^{}_{+})|^{2}\nonumber\\
&\times
|{\cal G}^{r}_{00}(\omega ^{}_{-})|^{2}
\sum_{\alpha,\alpha '={\rm L,R}}\Gamma^{}_{\alpha }(\omega^{}_{+})\Gamma^{}_{\alpha '}(\omega ^{}_{-})F^{}_{\alpha\alpha '}(\omega )\ .\label{NONCON}
\end{align}
This result, in conjunction with Eqs.  (\ref{IQP}) and (\ref{ELA}), leads to Eq. (\ref{DOTEP}). Finally, the net energy current carried by the electrons [see Eq. (\ref{IQ})] is obtained by using Eq. (\ref{SEQ}) in Eq. (\ref{ELA}) (and the corresponding equation for $\dot{E}_{\rm R}$). This yields Eq. (\ref{NEDOT}).

In the case of the ``floating"  molecule, which is not coupled to a heat bath of its own, it is straightforward to show [using the Hamiltonian (\ref{FH})]  that the rate of change of the phonon population \cite{WE} is given  by minus the right-hand side of Eq. (\ref{NONCON}) divided by $\omega_{0}$; i.e.,
\begin{align}
\int\frac{d\omega }{2\pi}\omega^{}_{}[{\cal I}^{\rm inel}_{\rm L}(\omega )+
{\cal I}^{\rm inel}_{\rm R}(\omega )]+\omega^{}_{0} \frac{dN}{dt} =0\ ,\label{N2}
\end{align}
yielding Eq. (\ref{ECon}) for the energy conservation, with $\dot{E}_{\rm P}=\omega_{0} dN/dt$.  However, the phonon population of a  floating molecule  will arrange itself according to the chemical potentials and the temperatures of the electronic reservoirs. \cite{WE}  Consequently at steady-state $dN/dt$ will vanish, implying that
$\dot{E}_{\rm L}+\dot{E}_{\rm R}=0$, such that the energy current of the electrons is conserved.

Finally we estimate  the rate of decay of the vibration population due to the coupling with the electrons in the leads, when the latter are at thermal equilibrium.  In diagrammatic language that rate is given by dressing the ``phonon" line with an electron bubble, i.e.
\begin{align}
&-\frac{dN}{dt}=\gamma^{2}\int\frac{d\omega}{2\pi}\frac{\Gamma (\omega^{}_{+})}{2}|{\cal G}_{00}^{a}(\omega^{}_{+})|^{2}\frac{\Gamma (\omega^{}_{-})}{2}|{\cal G}_{00}^{a}(\omega^{}_{-})|^{2}\nonumber\\
&\times\Bigl (Nf(\omega^{}_{-})[1-f(\omega^{}_{+})]-(1+N)f(\omega^{}_{+})[1-f(\omega^{}_{-})]\Bigr )\ .
\label{DN1}
\end{align}
At zero temperature the last factor in the integrand limits it to the range $|\omega -\mu|\leq\omega_{0}/2$, leading to the rate 
\begin{equation}
(\omega_{0}/[2\pi])(\gamma/\Gamma)^{2}
\label{phe}
\end{equation}  
when $\Gamma\gg\omega_{0}$, as  mentioned in Sec. \ref{INTRO}. In general, the rate is a non-monotonic function of the ratio $\omega_{0}/\Gamma$, reaching a maximal value when these two energies are comparable.

The result of Eq.~(\ref{phe}) can be qualitatively obtained in a more elementary, but equivalent, fashion using third-order perturbation theory. Consider the decay of the excited vibrational mode with a $T=0$ Fermi gas on each lead (having  a density of states $\nu_0$). For simplicity we take the resonant case and assume $\Gamma \gg \omega_0$ to start with. For the decay with the left lead, the first intermediate state has  
an electron from that lead go into the molecule with an amplitude $V^*_{\rm L}$ and an effective energy denominator $\Gamma_{\rm L}/2$ (due to being on resonance), in the second intermediate state, the vibration is deeexcited (amplitude $\gamma$) and the electrons stay put.  For   $\Gamma \gg \omega_0$, the energy denominator is again approximately $\Gamma_{\rm L}/2$. In the final state, the electron goes back to the same lead,
with an energy $ \omega_0$ higher that the one it started from, and with  an amplitude  $V_{\rm L}$. The total amplitude for this process is        $ V^* _{\rm L} \gamma V_{\rm L}/       (\Gamma_{\rm L} /2)^2$. Finally, taking the absolute square of the amplitude and multiplying by $2 \pi \nu_0$,  we get the golden-rule rate for this decay. Multiplying by the number of such processes,
$\nu_0  \omega_0$, and summing over the (equivalent) leads, we get Eq.~(\ref{phe}) in order of magnitude. Clearly, in the opposite case $\Gamma \ll \omega_0$, one $\Gamma$ in the denominator is replaced by $\omega_0$.
A surprising feature of this result, which must be pointed out, is its decrease with $\Gamma$, the rate to get into/from the molecule from/into the leads. This is quite counterintuitive, but it is what the quantum-mechanical calculation tells us! Formally, this is due to Green function at the resonance having  $i \Gamma / 2$ as its denominator, meaning physically that the width of the resonance sets the scale for the  ``closest approach" to it.

\begin{acknowledgments}
We thank Achim Rosch and Peter W\"olfle for illuminating discussions.
This work was supported by the German Federal Ministry of
Education and Research (BMBF) within the framework of the
German-Israeli project cooperation (DIP), by the US-Israel
Binational Science Foundation (BSF), by the Israel Science
Foundation (ISF) and by its Converging Technologies Program.
We thank the (anonymous) referee for pointing out the issues of the 
vibration relaxation by coupling to the electrons of the leads and the resulting 
thermal contact to a substrate. 
\end{acknowledgments}


\begin{thebibliography}{999}
\bibitem{COM1}
The use of this identity is justified by the assumption that the reservoirs supplying the particles are kept at thermal equilibrium.
\end{thebibliography}
\end{document}